\documentclass[aps,pra,showpacs]{revtex4}

\usepackage{graphicx}
\usepackage{dcolumn}
\usepackage{bm}
\usepackage{longtable}

\begin{document}

\title{Experimental evaluation of the non-classical relation between measurement errors using entangled photon pairs as a probe}

\author{Masataka Iinuma}
 \email{iinuma@hiroshima-u.ac.jp} 
 \homepage{http://home.hiroshima-u.ac.jp/qfg/qfg/index.html}
\affiliation{
 Graduate School of Advanced Sciences of Matter, Hiroshima University, 
 1-3-1 Kagamiyama, Higashi-Hiroshima, 739-8530, Japan
}%
\author{Yutaro Suzuki}
 \affiliation{Department of Physics, Graduate school of Science, Kyoto University, Kyoto, 606-8502, Japan}
 \altaffiliation{Present address: Magellan Systems Japan, Inc., Amagasaki Research Incubation Center 315, 
7-1-3, Doicho, Amagasaki, Hyogo, 660-0083, Japan}%
\author{Masayuki Nakano}
 \altaffiliation{Present address: NTTDATA CUSTOMER SERVICE Corporation, Sumitomo Fudosan Toyosu Building 4F, 9-6, Edagawa 1-choume, Koto-ku, Tokyo, 135-0051, Japan}%
\author{Holger F. Hofmann}
\affiliation{
 Graduate School of Advanced Sciences of Matter, Hiroshima University, 
 1-3-1 Kagamiyama, Higashi-Hiroshima, 739-8530, Japan
}%

\begin{abstract}
We have experimentally evaluated the non-classical relation between measurement errors in a joint measurement of two non-commuting polarizations by using entangled photon pairs as a probe. The joint measurement was realized by filtering polarization directions that are sensitive to both of the two target polarizations and the same filtering procedures were applied independently to the two photons of an entangled pair. Since the statistically independent measurement errors of two identical measurements performed on entangled pairs will reduce an overall visibility of the correlation by the square of the local visibilities, the squared visibilities of the local measurements can be obtained directly from the measurement data. We apply this method to determine the visibility of the correlation between the products of the two non-commuting polarizations, a characteristic of the measurement errors that can not be obtained locally since there is no self-adjoint operator that describes this product of measurement outcomes as a single photon observable. The experimental results clearly show that the square of the product visibility is negative, indicating the non-classical nature of the statistical relation between the measurement errors of non-commuting polarizations. 
\end{abstract}

\pacs{
03.65.Ta, 
42.50.Xa  
}

\maketitle

\section{Introduction}

In recent years, significant efforts have been made to better understand the role of uncertainties in quantum measurements \cite{Ozawa2003, Hall2004, Busch2007, Lund2010, Watanabe2011, Fujikawa2012, Erhart2012, Rozema2012, Branciard2013, Busch2013, Baek2013, Weston2013, Sulyok2013, Buscemi2014, Busch2014, Ringbauer2014, Kaneda2014, Zhou2016, Ma2016, Sulyok2017, Rodriguez2018}. Specifically, new methods have been developed to analyze quantum statistics based on joint measurements of non-commuting observables \cite{Lundeen2012, Vallone2016, Suzuki2016, Lundeen2016}. A major difficulty in the field is the problem of how to estimate the value of a physical property in the absence of precise measurement results \cite{Suzuki2012, Busch2013, Busch2014, Iinuma2016, Nii2017}. Without such an estimate, it is impossible to identify the value of a measurement error in an individual measurement. It is therefore important to consider the experimental procedures by which measurement errors can be evaluated in more detail. 

The only uncontroversial procedure for the evaluation of measurement errors is the use of an eigenstate input. It is then possible to evaluate the measurement errors for each property by comparing the input eigenvalue with the measurement result, confirming state independent uncertainty limits such as the ones given in \cite{Englert1996} for visibilities of two level systems. However, the error statistics observed with specific eigenstates do not indicate how measurement errors in a property $\hat{X}$ relate to measurement errors in a different non-commuting property $\hat{Y}$. As a consequence, it is difficult to judge whether the correlations between non-commuting properties that can be observed in uncertainty limited joint measurements of these two properties actually represent intrinsic correlations of the properties in the quantum state or merely result from correlations between the measurement errors \cite{Busch2013, Iinuma2016}. To solve this problem, we would need to identify unambiguous correlations between non-commuting properties in a well-defined input state. As shown in \cite{Kino2015}, this can be done by introducing entangled pairs as input, since such pairs are characterized by precise correlations between $\hat{X}_{1}$ and $\hat{X}_{2}$, and between $\hat{Y}_{1}$ and $\hat{Y}_{2}$. It is then possible to identify the statistical errors in the observation of the product value of $\hat{X}$ and $\hat{Y}$ from the observed changes in the correlations between the local products. Applied to two-level systems, the measurement statistics show a negative squared value for the visibility of the product of the non-commuting properties, a result that is qualitatively different  from any conventional statistical model of measurement errors. 

In the present paper, we realize the experiment proposed in \cite{Kino2015} using a source of entangled photons and a joint measurement composed of four carefully selected polarizer settings. The target properties are the horizontal (H) or vertical (V) polarization directions represented by the Pauli operator $\hat{X}$ and the positive (P) and negative (M) diagonal polarizations represented by the operator $\hat{Y}$. We confirm that the entangled input state shows very high levels of anti-correlations between $\hat{X}_{1}$ and $\hat{X}_{2}$, and between $\hat{Y}_{1}$ and $\hat{Y}_{2}$. These correlations are sufficient to identify a minimal correlation between the products of the measurement outcomes $X$ and $Y$. If the ability to observe the correct product value in a local measurement is represented by the product visibility $C$, the squared value $C^2$ determines the reduction of the input correlation between the actual product values in the entangled probe state. Our experimental results clearly show that the measurement errors invert the sign of the input correlation, indicating a negative value of $C^2$. We therefore conclude that the local value of $C$ must be imaginary, corresponding to a conversion of imaginary joint probabilities in possible input states into the real valued distribution observed in experiments. 

The rest of the paper is organized as follows. In Sec. \ref{sec:errors} we introduce measurement visibilities to describe the measurement errors of joint measurements in two-level systems and apply the description to two simultaneous local measurements performed on entangled pairs to show how the local visibilities can be evaluated from the experimentally observed correlations. In Sec. \ref{sec:exp}, the method of realizing a joint measurement of non-commuting polarizations using polarization filtering is explained and the experimental setup for the generation and measurement of polarization entangled photon pairs is described. In Sec. \ref{sec:results}, the experimental results are presented and the negative value of the squared product visibility is confirmed. Sec. \ref{sec:conclusions} concludes the paper.

\section{Evaluation of the visibilities of joint measurements on two-level systems using entangled pairs}
\label{sec:errors}
\begin{figure}[ht]
\includegraphics[width=100mm]{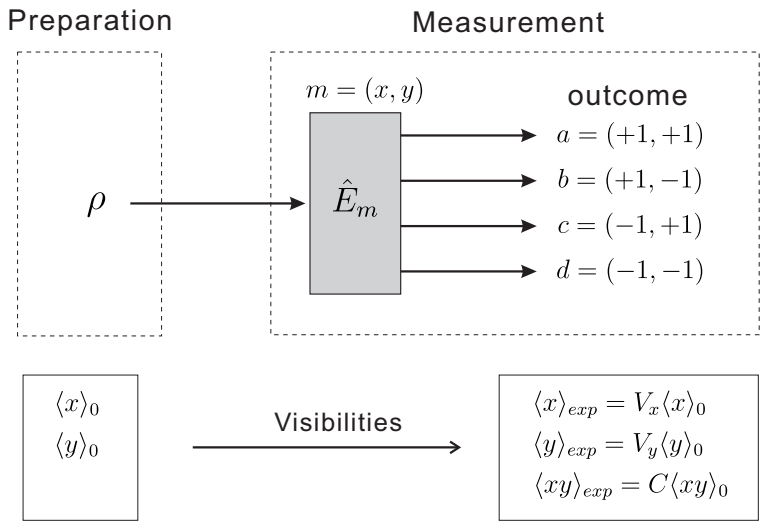}
\caption{Joint measurement of $\hat{X}$ and $\hat{Y}$. The $m$ represents a pair of two outcomes $(x, y)$ and can take following four possibilities, $a=(+1, +1)$, $b=(+1, -1)$, $c=(-1, +1)$, and $d=(-1, -1)$. The joint measurement provides three experimentally-obtained averages, $\langle x \rangle_{exp}$, $\langle y \rangle_{exp}$, and $\langle xy \rangle_{exp}$, which are related to the expectation values in the initial state, $\langle x \rangle_{0}$, $\langle y \rangle_{0}$, and the hypothetical initial expectation value $\langle xy \rangle_{0}$, through each visibility, which corresponds to the measurement error.}
\label{fig:joint}
\end{figure}

We consider joint measurements of two non-commuting properties $\hat{X}$ and $\hat{Y}$ of a two-level system, corresponding to two orthogonal Bloch vector directions. As shown in Fig. \ref{fig:joint}, the measurement results in one of four combinations of outcomes of $x=\pm 1$ and $y=\pm 1$.  Due to the uncertainty limit of the joint measurement, the expectation values of the outcomes $\langle x \rangle_{exp}$ and $\langle y \rangle_{exp}$ will be lower than the corresponding expectation values $\langle x \rangle_{0}$ and $\langle y \rangle_{0}$ of the input state. The resolution of the measurement can be characterized by the visibilities $V_{x}$ and $V_{y}$ defined as the ratios of the initial expectation values and the average value of the measurement outcomes,
\begin{eqnarray}
V_{x} &=& \frac{\langle x \rangle_{exp}}{\langle x \rangle_{0}}
\nonumber \\ 
V_{y} &=& \frac{\langle y \rangle_{exp}}{\langle y \rangle_{0}}. 
\label{eqn:defxy}
\end{eqnarray}
In addition to the expectation values of the two individual outcomes $x$ and $y$, a joint measurement also gives us information about the relation between $\hat{X}$ and $\hat{Y}$ in the form of the expectation value of the product, $\langle xy \rangle_{exp}$. This is particularly important because non-commutativity makes it impossible to define a corresponding expectation value for the local input state $\hat{\rho}$. In the following, we will solve this problem by using entangled photon pairs as a probe \cite{Kino2015}. For this purpose, it is useful to introduce a third visibility for the products of  $\hat{X}$ and $\hat{Y}$ based on the ratio of a hypothetical initial expectation value of the product $\langle xy \rangle_{0}$ and the average of the outcomes $\langle xy \rangle_{exp}$,
\begin{equation}
C = \frac{\langle xy \rangle_{exp}}{\langle xy \rangle_{0}}.
\label{eqn:defC}
\end{equation}
This definition does not provide a simple procedure for the experimental evaluation of $C$ due to the absence of a definition for the initial expectation value of a given local input state $\hat{\rho}$. Nevertheless it can be applied to independent measurements on entangled pairs if we can identify the correlation between the products of $\hat{X}$ and $\hat{Y}$ for the entangled input state. As shown in \cite{Kino2015}, this is indeed possible for Bell state inputs, since these states define precise correlations between both $\hat{X}$ and $\hat{Y}$, with well-defined consequences for the correlation of the products of $\hat{X}$ and $\hat{Y}$. Specifically, for a pair of two level systems the non-local products $\hat{X}_{1}\hat{X}_{2}$ and $\hat{Y}_{1}\hat{Y}_{2}$ commute with each other, so that it is possible to determine their joint probabilities $p(x_{1}x_{2}, y_{1}y_{2})$ and the expectation value of their product $\langle x_{1}x_{2}y_{1}y_{2} \rangle$, which is the same as the expectation value of the product of the four local quantities $x_{1}$, $y_{1}$, $x_{2}$, and $y_{2}$. Therefore, it is possible to define the initial expectation value $\langle (x_{1}y_{1}) (x_{2}y_{2}) \rangle_{0}$ of the entangled Bell state using the expectation value $\langle x_{1}x_{2}y_{1}y_{2} \rangle$ of a self-adjoint operator, even though this expectation value cannot be observed directly when using only local measurements. If we experimentally obtain the average value $\langle (x_{1}y_{1})(x_{2}y_{2}) \rangle_{exp}$ of the same correlation of products by using two joint measurements of the local non-commuting observables, their ratio with the known initial value of the Bell state gives the overall visibility for the correlation measurement, which can be factorized into two local visibilities, $C_{1}$ and $C_{2}$. Assuming that the two joint measurements are nearly identical, the square root of this ratio determines the local visibility $C=C_{1}=C_{2}$.  

\begin{figure}[ht]
\includegraphics[width=100mm]{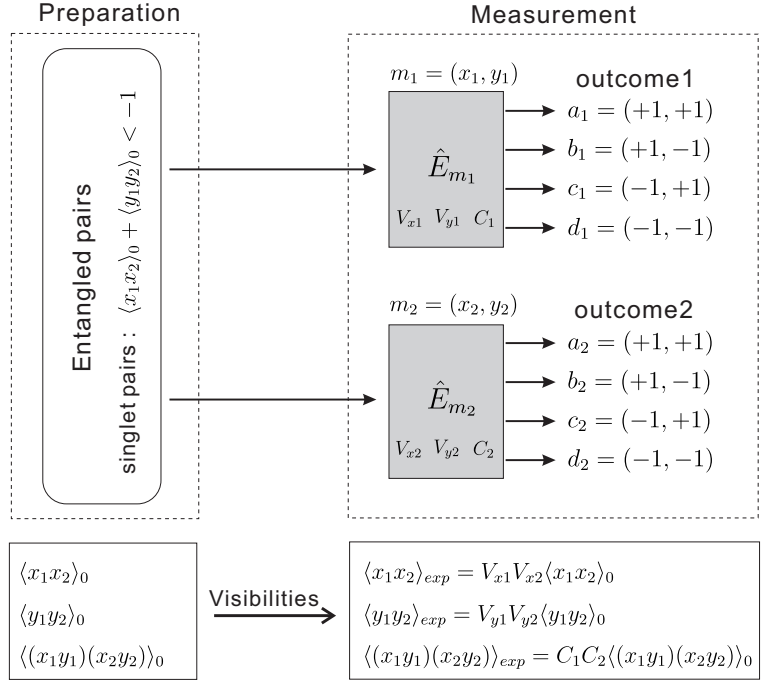}
\caption{Correlation measurement of the entangled pair with two local joint measurements. We can obtain three experimental correlations, $\langle x_{1}x_{2} \rangle_{exp}$, $\langle y_{1}y_{2} \rangle_{exp}$, and $\langle (x_{1}y_{1})(x_{2}y_{2}) \rangle_{exp}$ from two combination outcomes in the two joint measurements, which correspond to three initial correlations of the entangled input pairs. The overall visibilities of the two joint measurements can be expressed by a product of the two local visibilities, since the visibilities represent measurement errors that occur independently in the two local measurements.}
\label{fig:two_joint}
\end{figure}

Fig. \ref{fig:two_joint} illustrates the evaluation of visibilities by correlation measurements of the entangled pairs using two local joint measurements. The two joint measurements result in a combination of two outcomes, $m_{1}=(x_{1}, y_{1})$ and $m_{2}=(x_{2}, y_{2})$, for the measurements $\hat{E}_{m_{1}}$ and $\hat{E}_{m_{2}}$, respectively. From the statistics of the measurement outcomes, we can obtain three experimental averages, $\langle x_{1}x_{2} \rangle_{exp}$, $\langle y_{1}y_{2} \rangle_{exp}$, and $\langle (x_{1}y_{1})(x_{2}y_{2}) \rangle_{exp}$, which are related to the initial expectation values, $\langle x_{1}x_{2} \rangle_{0}$, $\langle y_{1}y_{2} \rangle_{0}$, and $\langle (x_{1}y_{1})(x_{2}y_{2}) \rangle_{0}$, as follows, 
\begin{eqnarray}
\langle x_{1}x_{2} \rangle_{exp} & = & V_{x1}V_{x2}\langle x_{1}x_{2} \rangle_{0} \nonumber \\[4pt]
\langle y_{1}y_{2} \rangle_{exp} & = & V_{y1}V_{y2}\langle y_{1}y_{2} \rangle_{0} \nonumber \\[4pt]
\langle (x_{1}y_{1})(x_{2}y_{2}) \rangle_{exp} & = & C_{1}C_{2}\langle (x_{1}y_{1})(x_{2}y_{2}) \rangle_{0}, 
\label{eqn:visibilities}
\end{eqnarray}
where the overall visibilities of the two-measurement system can each be expressed as products of two local visibilities, $V_{x1}V_{x2}$, $V_{y1}V_{y2}$, and $C_{1}C_{2}$ since the local measurements are independent of each other.  

The experimental averages in Eq.(\ref{eqn:visibilities}) can be efficiently evaluated by classifying the measurement outcomes according to the errors in the correlations between the local outcomes, where the correct values are known from the Bell state inputs. In the following, we use $E_{ij}$ to denote the average probability of outcomes with a specific error pattern, where $i=0$ indicates the correct correlation in $x$ and $j=0$ indicates the correct correlation in $y$, while values of $i=1$ and $j=1$ indicate the corresponding errors. Table \ref{tab:corr} shows which measurement outcomes contribute to which probability average $E_{ij}$ in the singlet Bell state input.
\begin{table}[h]
\begin{ruledtabular}
\begin{tabular}{c|cccc} 
                    & \multicolumn{4}{c}{outcome2 $(x_{2}, y_{2})$} \\ 
outcome1 $(x_{1}, y_{1})$  & $(+1,+1)$ & $(+1,-1)$ & $(-1,+1)$ & $(-1,-1)$  \\ \cline{2-5}
$(+1,+1)$         & $E_{11}$ & $E_{10}$ & $E_{01}$ & $E_{00}$ \\ 
$(+1,-1)$         & $E_{10}$ & $E_{11}$ & $E_{00}$ & $E_{01}$ \\ 
$(-1,+1)$         & $E_{01}$ & $E_{00}$ & $E_{11}$ & $E_{10}$ \\
$(-1,-1)$         & $E_{00}$ & $E_{01}$ & $E_{10}$ & $E_{11}$ \\ 
\end{tabular}
\caption{Contribution of measurement outcomes to probability averages based on the possible correlation error combinations in the singlet Bell state input.}
\label{tab:corr}
\end{ruledtabular}
\end{table}

The experimental averages can then be expressed by only four probability averages,
\begin{eqnarray}
\langle x_{1}x_{2} \rangle_{exp}
& = & 4 ( -E_{00}-E_{01}+E_{10}+E_{11} ) \nonumber \\
\langle y_{1}y_{2} \rangle_{exp}
& = & 4 ( -E_{00}-E_{10}+E_{01}+E_{11} ) \nonumber \\
\langle (x_{1}y_{1})(x_{2}y_{2}) \rangle_{exp} 
& = & 4 ( +E_{00}+E_{11}-E_{01}-E_{10} ), 
\label{eqn:ave_theory}
\end{eqnarray}
where we use $x_{1}x_{2}=-1$ and $y_{1}y_{2}=-1$ for the correct correlation and $x_{1}x_{2}=+1$ and $y_{1}y_{2}=+1$ for the incorrect one, corresponding to the correlations of the Bell state we use for the experiment.  

In general, it is necessary to evaluate the initial expectation value $\langle (x_{1}y_{1}) (x_{2}y_{2}) \rangle_{0}$ experimentally from the actual entangled state used as an input. Since we want to use as little theoretical assumptions as possible, we obtain an estimate of the minimal and maximal values from the correlations observed between local measurements of $\hat{X}$ and of $\hat{Y}$ for the experimentally generated entangled pairs. Since we generate entangled photons with opposite polarizations, we can expect to find correlations $\langle x_{1}x_{2} \rangle_{0}$ and $\langle y_{1}y_{2} \rangle_{0}$ close to minus one. However, the occasional emission of photons with equal polarizations reduces this value, and the maximal reduction of the product correlation is equal to the sum of the individual reductions, resulting in a lower estimate of the product correlation of
\begin{equation}
\langle x_{1}x_{2}y_{1}y_{2} \rangle_{0} \ge 1-(1+\langle x_{1}x_{2} \rangle_{0})-(1+\langle y_{1}y_{2} \rangle_{0}).
\end{equation}
For use in Eq.(\ref{eqn:visibilities}), it is important that the visibility cannot be reduced to zero or to negative values, since this would make a reliable reconstruction of the visibilities impossible. We therefore find that the minimal correlations needed at the input are given by
\begin{equation}
\langle x_{1}x_{2} \rangle + \langle y_{1}y_{2} \rangle < -1, 
\label{eqn:cond_ent}
\end{equation}  
Interestingly, this condition indicates that the input photons must be entangled to ensure that we can identify a non-vanishing correlation between the products of $\hat{X}$ and $\hat{Y}$.  

Once we are sure about the sign of the correlation, we also need to consider the upper bound of the correlation. Assuming that the correlations $\langle x_{1}x_{2} \rangle_{0}$ and $\langle y_{1}y_{2} \rangle_{0}$ are nearly equal, it is always possible that the correlation of the products is close to one, and this is the worst case scenario for the evaluation of measurement visibilities based on Eq.(\ref{eqn:visibilities}). We should therefore consider the whole range of possible product correlation values,
\begin{equation}
(-\langle x_{1}x_{2} \rangle_{0}-\langle y_{1}y_{2} \rangle_{0} -1) \le \langle (x_{1}y_{1})(x_{2}y_{2}) \rangle_{0} \le 1. 
\label{eqn:init_ave_limit}
\end{equation}
We can now use the experimental probability averages $E_{ij}$ to evaluate the visibilities of the local measurements. If the two joint measurements are identical, the squares of the local visibilities $V_{x}=V_{x1}=V_{x2}$ and $V_{y}=V_{y1}=V_{y2}$ can be obtained as follows, 
\begin{eqnarray}
V_{x}^{2} & = & \frac{4 ( E_{00}+E_{01}-E_{10}-E_{11} )}{ |\langle x_{1}x_{2} \rangle_{0}| } \nonumber \\
V_{y}^{2} & = & \frac{4 ( E_{00}+E_{10}-E_{01}-E_{11} )}{ |\langle y_{1}y_{2} \rangle_{0}| } 
\label{eqn:vis_XY}
\end{eqnarray}
In addition, we obtain an upper and a lower bound for the product visibility $C=C_{1}=C_{2}$,
\begin{equation}
4 ( E_{00}+E_{11}-E_{01}-E_{10} ) \le C^{2} \le \frac{4 ( E_{00}+E_{11}-E_{01}-E_{10} )}{(-\langle x_{1}x_{2} \rangle_{0}-\langle y_{1}y_{2} \rangle_{0} -1)}. 
\label{eqn:vis_C}
\end{equation}
These equations show that the squares of the local visibilities can be determined from the output statistics of the joint measurements using the distribution of correlation errors described by the probability averages $E_{ij}$. Importantly, the possibility of estimating the value of $C^2$ using Eq.(\ref{eqn:vis_C}) depends on the presence of entanglement in the input as shown by Eq.(\ref{eqn:cond_ent}). If the input correlations are sufficiently high, the experimentally observed product correlation is proportional to $C^2$, with only a small range of possible ratios between the experimental result and $C^2$. 

\section{Experimental setup}
\label{sec:exp}
\subsection{Polarization measurements}
For the experiment, we realized a joint measurement of two non-commuting polarization, $\hat{S}_{HV}=\hat{X}$ representing horizontal(H) and vertical(V) polarization and $\hat{S}_{PM}=\hat{Y}$ representing diagonal linear polarizations with $45^{\circ}$(P) and $135^{\circ}$(M), respectively. The measurement was performed by selecting four polarization directions which were oriented along intermediate angles between the H/V and the P/M polarization. In addition, a phase shift between H and V polarization was introduced to the outcomes $(H,P)=a$ and $(V,M)=d$, and the opposite phase shift was introduced to $(H,M)=b$ and $(V,P)=c$. The polarization directions can be characterized in terms of their orientations on the Bloch sphere, where the $\langle \hat{Z} \rangle$ direction corresponds to the circular polarization component $\hat{S}_{RL}=\hat{Z}$. The Bloch vector directions can be given by the spherical coordinates $\theta_{B}$ and $\phi_{B}$. It is convenient to define these angles with respect to the $\langle \hat{X} \rangle$ axis, such that $\theta_{B}$ is the angle between the $\langle \hat{X} \rangle$ axis and the Bloch vector, and $\phi_{B}$ is the rotation angle around the $\langle \hat{X} \rangle$ axis, starting from the positive $\langle \hat{Y} \rangle$ axis and rotating towards the $\langle \hat{Z} \rangle$ axis as $\phi_{B}$ increases. 

In general, the four outcomes represent results of $m=(s_{HV}, s_{PM})$. For simplicity, we have labeled the four outcomes $a=(+1,+1)$, $b=(+1,-1)$, $c=(-1,+1)$, and $d=(-1,-1)$. To realize a valid joint measurement, the sum of the Bloch vectors for $a$ and $b$ have to point in the positive $\langle \hat{X} \rangle$ direction (towards $H$) and the sum of the Bloch vectors for $c$ and $d$ have to point in the negative $\langle \hat{X} \rangle$ direction (towards $V$). Likewise, the sums of $a$ and $c$ have to point towards the positive $\langle \hat{Y} \rangle$ direction and the sums of $b$ and $d$ have to point towards the negative $\langle \hat{Y} \rangle$ direction. In all cases, the visibility is determined by the angle between the corresponding axis and the Bloch vectors of the measurement outcomes. In the following, we choose a setting where the theoretically predicted visibilities are given by $V_x = |\cos \theta_{B}|=1/\sqrt{2}$ and $V_y = |\sin \theta_{B} \cos \phi_{B}|=1/2$. In this case, the additional phase shift between the horizontal and vertical polarization components tilts the Bloch vectors towards the $\langle \hat{Z} \rangle$ axis, which means that the correlation between the measurement outcomes is sensitive to the circular polarization component $\hat{S}_{RL}=\hat{Z}$. As mentioned in \cite{Kino2015,Suzuki2016}, this is consistent with the observation that the operator product of $\hat{X}$ and $\hat{Y}$ is given by $\hat{X}\hat{Y}=i\hat{Z}$. 

The actual measurement was realized by a sequence of a quarter-wave-plate (QWP), a half-wave-plate(HWP) and a polarization beam splitter(PBS). To select the polarization directions, we determined the necessary angles for the QWP and the HWP. The angles $\theta_{H}$ of the HWP and $\theta_{Q}$ of the QWP used in the experiment are shown in Tab. \ref{tab:angle}, together with the spherical coordinates $\theta_{B}$ and $\phi_{B}$ describing the Bloch vector direction with respect to the $\langle \hat{X} \rangle$ axis. Note that the visibilities realized by these angles are $V_x=1/\sqrt{2}$ and $V_y=1/2$, so that the measurement is more sensitive to the HV polarization than to the PM polarization. 

\begin{table}[ht]
\begin{ruledtabular}
\begin{tabular}{ccccc} 
 Polarization direction  & $\theta_{B}$ & $\phi_{B}$ & $\theta_{H}$  & $\theta_{Q}$ \\ \colrule
$a=(+1,+1)$ & $45^\circ$ & $45^\circ$ & $16.32^\circ$ & $17.63^\circ$ \\ 
$b=(+1,-1)$ & $45^\circ$ & $-135^\circ$ & $-16.32^\circ $ & $-17.63^\circ$ \\ 
$c=(-1,+1)$ & $135^\circ$ & $-45^\circ$ & $28.68^\circ$ & $72.36^\circ$ \\ 
$d=(-1,-1)$ & $135^\circ$ & $135^\circ$ & $61.32^\circ$ & $107.63^\circ$ \\ 
\end{tabular}
\end{ruledtabular}
\caption{Angles at which the wave plates were set to realize the joint measurement of non-commuting polarizations. $\theta_{B}$ and $\phi_{B}$ are the spherical coordinates of the polarization direction on the Bloch sphere with respect to the $\langle \hat{X} \rangle$ axis, and $\theta_{H}$ and $\theta_{Q}$ are the angles at which the HWP and the QWP were set, respectively.}
\label{tab:angle}
\end{table}

\subsection{Entanglement source and photon detection}
\begin{figure}[h]
\centerline{\includegraphics[width=70mm]{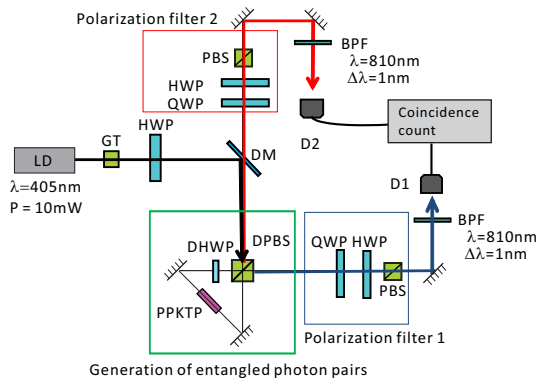}} 
\caption{Experimental setup for entanglement generation and polarization measurements. Pump light is generated by a laser diode (LD) with a wavelength of 405 nm and an output power of 10 mW. The pump light passes through a Glan-Thompson(GT) prism and a half-wave-plate(HWP) before being reflected at a dichroic mirror (DM) and entering the triangle Sagnac-type interferometer through the double polarization beam splitter(DPBS), where it pumped a periodically poled KTP(PPKTP) crystal from both sides, generating counter propagating photon pairs. The double half-wave-plate (DHWP) rotates the polarization of both 405 nm pump light and 810 nm down-converted photons, generating the intended polarization entanglement by interference at the DPBS. Polarization filters for the two output ports are constructed by a sequence of quarter-wave-plate (QWP) and half-wave-plate (HWP) rotated to the appropriate angles and followed by a polarization beam splitter (PBS). After passing through a band pass filter (BPF) with a bandwidth of about 1 nm, the photons were detected by the photon counting modules (Excelitas SPCM-AQS04) D1 and D2 and coincidence events were recorded based on the electronic signals from the detectors. }
\label{fig:setup}
\end{figure}

The polarization-entangled photon pairs were generated using a periodically poled KTP(PPKTP) crystal(Type II) inserted into a triangle Sagnac-type interferometer \cite{Kim2006, Fedrizzi2007, Jin2014}. The crystal had a dimension of 1 mm $\times$ 1 mm $\times$ 10 mm and was pumped by a laser diode(LD) with a wavelength of 405 nm and a power of 10 mW at a fiber output. Fig. \ref{fig:setup} shows the complete setup. The intended polarization entanglement was obtained by 
separating the pump beam into two circulating paths at the double polarization beam splitter (DPBS) serving as both the input port for the pump light at 405 nm and the output port for the down-converted photons at 810 nm. Coherent superpositions of photon pairs circulating clockwise or anti-clockwise were generated, where the double half-wave-plate (DHWP) adjusts the polarization of the clockwise propagating pair to produce the desired entangled polarization state at the output by interference between clockwise or anti-clockwise circulating pairs at the DPBS. 

After passing through the polarization filters described above, the photon pairs were detected using two single photon counting modules(Excelitas SPCM-AQS04) in each output path. After the electric TTL pulses from the detector were converted to electric NIM pulses with a width of 8 ns, coincidence pulses were generated with the NIM module and re-converted to TTL pulses for data acquisition. The maximum coincidence count rate in anti-correlation was about 3000 s$^{-1}$. For all settings of the polarization filters, photon pairs were counted over a period of 10 s to obtain a normalized value for the count rate. 

To evaluate the polarization correlations of the entangled photon pairs, we first measured three correlations between two local polarizations on the $\hat{S}_{HV}$, on the $\hat{S}_{PM}$, and on the $\hat{S}_{RL}$, which represents Right-handed ($R$) and Left-handed ($L$) polarizations. The specific results were 
\begin{eqnarray}
\langle s_{HV1} s_{HV2} \rangle_{0} & = & -0.9551\pm0.0012 \nonumber \\ 
\langle s_{PM1} s_{PM2} \rangle_{0} & = & -0.8555\pm0.0022 \nonumber \\
\langle s_{RL1} s_{RL2} \rangle_{0} & = & -0.8556\pm0.0022,  
\label{eqn:init_value_exp}
\end{eqnarray}
where the condition given by Eq.(\ref{eqn:cond_ent}) is satisfied by a value of $-\langle s_{HV1} s_{HV2} \rangle_{0} - \langle s_{PM1} s_{PM2} \rangle_{0} = 1.8106 > 1$. This result also serves as an entanglement witness, demonstrating the entanglement of the photon pairs generated by our source. We can now use Eq. (\ref{eqn:init_ave_limit}) to obtain an estimate of the product correlation, 
\begin{equation}
0.811 < \langle (s_{HV1}s_{PM1})(s_{HV2}s_{PM2}) \rangle_{0} \le 1. 
\label{eqn:init_value_C_exp}
\end{equation}
As expected, the lower bound of the correlation is sufficiently close to one to provide reliable information about the product visibility $C$ when the joint measurement is applied to both photons of the entangled pairs.

\section{Results and discussion}
\label{sec:results}
\begin{figure}[h]
\centerline{\includegraphics[width=100mm]{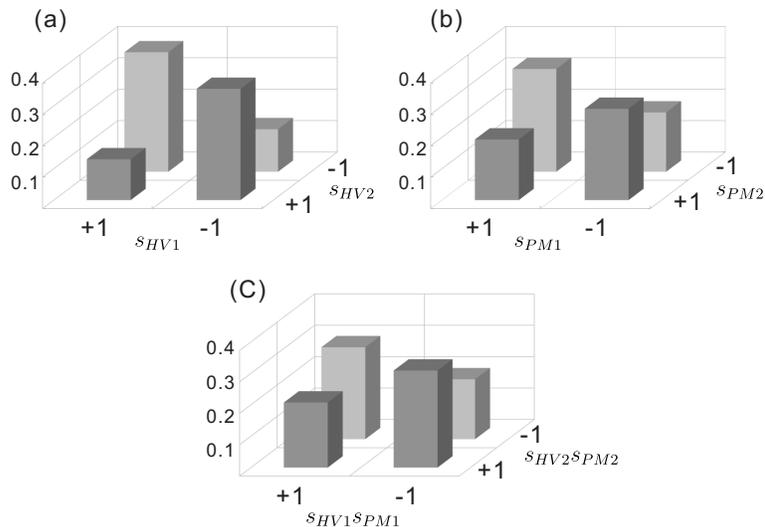}} 
\caption{Experimental results for the outcomes of (a) $s_{HV1}$ and $s_{HV2}$, of (b) $s_{PM1}$ and $s_{PM2}$, and of (c) the products $s_{HV1}s_{PM1}$ and $s_{HV2}s_{PM2}$. The results clearly show that opposite results are most likely in all three cases. This presents a paradoxical situation in the case of the product correlation shown in (c), since the input correlation for the entangled state was positive, suggesting a higher probability for equal outcomes in the input state.}
\label{fig:marginal}
\end{figure}

By setting the HWPs and the QWPs in two paths at the appropriate angles, we obtained the count rates of all 16 combinations of joint measurement outcomes $(s_{HV1}, s_{PM1})$ for photon 1 and joint measurement outcomes $(s_{HV2}, s_{PM2})$ for photon 2. The results are summarized in Tab. \ref{tab:result}. From these results, the average probabilities $E_{ij}$ associated with the different correlation errors can be evaluated, reducing the error statistics to four characteristic values of
\begin{eqnarray}
E_{00} &=& 0.0938 \pm 0.0004,
\nonumber \\
E_{01} &=& 0.0898 \pm 0.0004,
\nonumber \\
E_{10} &=& 0.0607 \pm 0.0004,
\nonumber \\
E_{11} &=& 0.00567 \pm 0.00094.
\end{eqnarray}
We can now use Eq.(\ref{eqn:ave_theory}) to determine the experimentally observed correlations,
\begin{eqnarray}
\langle s_{HV1}s_{HV2} \rangle_{exp} & = &  -0.469 \pm 0.003,\nonumber \\
\langle s_{PM1}s_{PM2} \rangle_{exp} & = & -0.236 \pm 0.003,\nonumber \\
\langle s_{HV1}s_{PM1}s_{HV2}s_{PM2} \rangle_{exp} & = & -0.204 \pm 0.003. 
\label{eqn:exp_corr}
\end{eqnarray}
To visualize these experimental results, Fig. \ref{fig:marginal} shows the bar graphs for the probabilities of $s_{HV1}$ and $s_{HV2}$ in Fig. \ref{fig:marginal} (a), for the probabilities of $s_{PM1}$ and $s_{PM2}$ in Fig. \ref{fig:marginal} (b), and for the product values $s_{HV1}s_{PM1}$ and $s_{HV2}s_{PM2}$ in Fig. \ref{fig:marginal} (c). In all three graphs, opposite results clearly dominate the probability distribution, as confirmed by the negative signs of the values in Eq.(\ref{eqn:exp_corr}). 

\begin{table}[h]
\begin{ruledtabular}
\begin{tabular}{c|cccc} 
outcome 1 & \multicolumn{4}{c}{outcome 2 $(s_{HV2}, s_{PM2})$} \\ 
$(s_{HV1}, s_{PM1})$  & $(+1,+1)$ & $(+1,-1)$ & $(-1,+1)$ & $(-1,-1)$ \\ \cline{2-5}
$(+1,+1)$ & $   967  $ & $ 8723 $ & $ 16658 $  & $ 17558 $ \\ 
$(+1,-1)$ & $ 10341 $ & $  834   $ & $ 12621  $ & $ 14248 $ \\ 
$(-1,+1)$ & $ 12521 $ & $ 16356 $ &  $  864   $ & $ 9934 $\\ 
$(-1,-1)$ & $ 13736 $ & $ 14248 $ & $ 9972 $ & $   996  $ \\ 
\end{tabular}
\end{ruledtabular}
\caption{Table of the number of coincidence counts observed for the 16 different measurement outcomes (polarization settings) within a fixed period of 10 s. The magnitude of statistical errors can be estimated by the square root of the number of counts registered.}
\label{tab:result}
\end{table}

It is now possible to determine the visibilities of the joint polarization measurement by comparing the experimentally observed correlations to the input correlations of the entangled photon pairs. For the individual polarization measurements, the results read
\begin{eqnarray}
V_{x}^{2} & = & 0.491 \pm 0.003 \nonumber \\
V_{y}^{2} & = & 0.276 \pm 0.004 
\label{eqn:results_vis_XY}
\end{eqnarray}
These results are reasonably close to the values expected from the Bloch vector directions that define the measurement. Specifically, the predicted values for the settings in Tab. \ref{tab:angle} are $V^{2}_{x}=\cos^2 \theta_{B}=0.5$ and $V^{2}_{y}=( \sin \theta_{B} \cos \phi_{B} )^2=0.25$. The slight deviation of the experimental results from these expected values may well be caused by small systematic errors in the settings of the waveplate angles, since the visibilities are rather sensitive to small changes in these angles. 

The main result is obtained for the visibility $C$ of the product $s_{HV}s_{PM}$. Here, the input correlation has a positive value between $0.811$ and $1$ as shown by Eq.(\ref{eqn:init_value_C_exp}), but the observed output correlation given in Eq.(\ref{eqn:exp_corr}) has a negative value of $-0.204 \pm 0.003$. As a consequence, we find that the experimental results indicate that the square of the local product visibility has a negative value bounded by
\begin{equation}
-0.252 \pm 0.004  <  C^{2}  \le -0.204 \pm 0.003.  
\label{eqn:result_vis_C}
\end{equation}
The negative value of the squared visibility suggests that the local visibility is imaginary, with a value of
\begin{equation}
0.452 \pm 0.003  \le  \pm i C  < 0.502 \pm 0.004.  
\label{eqn:imag_C}
\end{equation}
As mentioned above and in \cite{Kino2015}, the imaginary visibility seems to indicate that the operator product $\hat{X}\hat{Y}=i \hat{Z}$ might be a valid expression relating the products of $s_{HV}s_{PM}$ to the circular polarization of the photon through an imaginary factor of $i$.  Circular polarization thus appears as an imaginary value of the product of $\hat{X}$ and $\hat{Y}$, which can only appear in the measurement statistics after being converted into a real expectation value by an imaginary value of the product visibility $C$. In the present experiment, the $\langle \hat{Z} \rangle$ components of the Bloch vector describing the measurement outcomes was $1/2$, which means that we expect to find an imaginary visibility of $C=\pm 0.5 i$. The experimental result is therefore fully consistent with the theory presented in \cite{Kino2015}.

It may be interesting to consider whether there might be an alternative explanation of the measurement result, based on the assumption that the negative sign of the product correlation originates from differences in the measurement errors of the two measurements. In this case, the two local visibilities $C_{1}$ and $C_{2}$ would have to have opposite signs. At a visibility of about $\pm 0.5$, this would be a rather drastic difference between the two measurements, which is difficult to reconcile with the fact that both measurements were carried out in the same manner. The significance of the experimental result arises from the qualitative difference between the negative correlation of $\langle s_{HV1}s_{PM1}s_{HV2}s_{PM2} \rangle_{exp} = -0.204$ observed in the experiment, and the positive correlation of $\langle (s_{HV1}s_{PM1})(s_{HV2}s_{PM2}) \rangle_{0} > 0.811$ of the input pairs. The inversion of the sign is a characteristic signature of the quantum correlations between non-commuting observables that cannot be reproduced by any corresponding classical scenario.  

\section{Conclusions}
\label{sec:conclusions}
We have performed joint measurements of two non-commuting polarization components $\hat{S}_{HV}$ and $\hat{S}_{PM}$ on pairs of polarization-entangled photons. Since the correlations $\langle s_{HV1}s_{HV2} \rangle$ and $\langle s_{PM1}s_{PM2} \rangle$ were sufficiently close to their ideal values of $-1$, it was possible to determine a minimal positive value for the product correlation of $\langle (s_{HV1}s_{PM1})(s_{HV2}s_{PM2}) \rangle_{0} > 0.811$. It is then possible to estimate the squared visibility $C^2$ of the measurement of the product value $s_{HV} s_{PM}$ from the correlation of measurement outcomes in the joint measurements. 

As predicted by theory \cite{Kino2015}, the experimentally observed correlation of $\langle (s_{HV1}s_{PM1})(s_{HV2}s_{PM2}) \rangle_{exp} =- 0.204 \pm 0.003$ has a negative sign, which corresponds to a negative squared visibility of $C^2 \le - 0.204 \pm 0.003$. Our results thus demonstrate that the joint measurement statistics of entangled pairs can reveal non-classical correlations between the local measurement errors of the non-commuting observables, strongly suggesting that the imaginary part of operator products expressed by commutation relations corresponds to imaginary correlations in the quantum statistics. Quantum correlations in the measurement errors ensure that such imaginary parts in the joint quantum statistics of non-commuting observables appear as real-valued probabilities in the statistics of uncertainty limited quantum measurements. The experimental approach offered by uncertainty limited joint measurements of non-commuting observables may therefore provide us with a better practical understanding of how the physics described by Hilbert space states and operators actually works.  

\section*{Acknowledgments}
We are grateful to Geoff Pryde, Alessandro Fedrizzi, and  Rui-Bo Jin for their helpful suggestions regarding the construction of the entangled photon source. This work was supported by JSPS KAKENHI Grant Number 24540428. One of authors (Y.S.) was supported by Grant-in-Aid for JSPS Research Fellow Number JP14J05259.

\end{document}